\documentclass[
reprint,                
superscriptaddress,     
amsmath,                
amssymb,                
prab,                   
]{revtex4-2}            

\usepackage[utf8]{inputenc} 
\usepackage[T2A]{fontenc}   
\usepackage[russian,english]{babel} 

\usepackage{color,soul}     
\sethlcolor{yellow}         
\usepackage[autostyle]{csquotes} 

\usepackage{graphicx}       
\graphicspath{{figures/}}   

\usepackage{dcolumn}        
\usepackage{multirow}       
\usepackage{enumitem}       

\usepackage{bm}             
\usepackage{amsmath}        
\usepackage{mathtools}      
\usepackage{braket}         
\usepackage{dsfont}         

\usepackage[mathlines]{lineno} 

\newcommand{\kbri}[3]{\ket{#1}_{#3}\!\bra{#2}}

\newcommand{\iket}[2]{\ket{#1}_{#2}}
\newcommand{\ibra}[2]{\prescript{}{#2}{\bra{#1}}}

\newcommand{\dmspace}[1]{\mathcal{S}\left( #1 \right)}

\DeclareMathOperator{\tr}{Tr}

\newenvironment{protocol}
{
	\vspace{0.1em}
	\noindent\rule{\linewidth}{0.5pt}\vspace{-1em}\\
	\noindent\rule{\linewidth}{0.5pt} \vspace{-3em}\\
}
{ 
	\vspace{-2.5em} \leavevmode \\ 
	\noindent\rule{\linewidth}{0.5pt} \vspace{-2.25em} \\
	\noindent\rule{\linewidth}{0.5pt} \vspace{-2em} \\
}

\begin{document}

\preprint{APS/123-QED}

\title{Modulator-free transmitter for quantum key distribution\\
	in metropolitan area networks}

\author{Roman Shakhovoy}
\email{r.shakhovoy@goqrate.com}  
\affiliation{QRate, Moscow, Russia}
\affiliation{NTI Center for Quantum Communications, National University of Science and Technology MISIS, Moscow, Russia}

\author{Evgeniy Dedkov}
\affiliation{QRate, Moscow, Russia}

\author{Igor Kudryashov}
\affiliation{QRate, Moscow, Russia}

\date{\today}

\begin{abstract}
A positive economic effect from the implementation of quantum key distribution (QKD) technology can be achieved only with significant scaling, which involves the deployment of branched metropolitan area networks. The creation of QKD systems suitable for such networks is an important task for the coming years. This paper considers a method for preparing quantum states using pulsed optical injection, which can be used as a basis for a compact modulator-free transmitter ideally suited for QKD at typical distances within a city. Considering the relative proximity between nodes of a MAN, we suggest to abandon the decoy states, which, together with the proposed method of quantum state preparation, allows making the transmitter extremely simple. We report here the results of an experiment confirming the operating principle and provide a security analysis of the three-state decoy-free QKD protocol that can be implemented using such a device.
\end{abstract}

\maketitle

\section{Introduction}\label{sec:introduction}
The emergence and rapid development of quantum key distribution (QKD) networks around the world \cite{Elliot2003,Peev2009,Sasaki11,Zhang18} demonstrates not only the maturity of this technology, but also its readiness for widespread implementation. It can therefore be expected that in the near future QKD networks will become an integral part of the IT infrastructure. (We note, however, the criticism of some governments to the QKD \cite{NSA2023,NCSC2023,ANSSI2023} and the ongoing discussion in academic circles aimed at finding counterarguments to such criticism \cite{Renner2023debate,QCH2023}.) Judging by the experience of implementing classical telecommunication networks, a positive economic effect from the implementation of QKD networks can be achieved only with their significant scaling. For this, not only backbone networks are required, but branched metropolitan area QKD networks (QKD MANs) are demanded. Modern manufacturers of QKD systems, however, are focused primarily on the development of high-cost terminals for backbone nodes, the use of which in QKD MANs can hardly be considered economically justified. In classical telecommunication networks, e.\,g., there is a significant difference in the cost of equipment for wide (WANs), metropolitan (MANs), and local area networks (LANs), which, in fact, allows them to be effectively scaled. Thus, an important task for the coming years is the creation of cost-effective (inexpensive) QKD systems suitable for MANs.

One of the approaches to creating equipment for QKD MANs is continuous-variables QKD \cite{Ralph99}, which has a number of advantages over discrete-variables QKD. The main advantage is that single-photon detectors are not required in this case --- relatively cheap coherent receivers operating at room temperature \cite{Diamanti2015} are used instead. In addition, such an approach can potentially provide a higher key generation rate at short distances \cite{Wang2022,Pan2022}. However, encoding information in an infinite-dimensional Hilbert space has disadvantages. In particular, complex post-processing is required, including special error-correction codes \cite{Alleaume2021}. In addition, CV-QKD is sensitive to losses in the quantum channel and requires the presence of a common reference phase between the receiver (Bob) and the transmitter (Alice), which is a specific problem for QKD methods using coherent detection.

Another approach is to create a passive source \cite{Curty2010, Wang2023fully}. The advantage is that the optical circuit of the quantum transmitter does not contain modulators, so potential side channels introduced by active components are eliminated. Since optical modulators and the hardware required to operate them constitute a significant part of the quantum transmitter's cost, this approach can potentially make the system cheaper. Another important advantage of this approach is that passive state preparation does not require a random number generator, which further simplifies the system. A significant disadvantage, however, is the need to postselect the quantum states. This means that Alice must measure her states before sending them to know what she is sending, and only a portion of the prepared states is used to generate the key, which negatively affects the key rate. In addition, the need for post-selection imposes additional requirements on the transmitter functionality, preventing it from being simple enough.

The third approach also consists of using a transmitter without electro-optical modulators, however, unlike the passive-source approach, here the states are prepared actively using pulsed optical injection \cite{Yuan2016,Roberts2018,Paraiso2019,Lo2023simplified}. Such a technique was first proposed in \cite{Yuan2016}, analyzed theoretically using the rate equation method in \cite{Shakhovoy2021direct} and has already demonstrated its effectiveness in practice \cite{Roberts2018,Paraiso2019}. The approach potentially allows for simplification (and significant reduction in cost) of the quantum transmitter, and also opens up broad opportunities for further miniaturization while maintaining all the advantages of QKD on discrete variables.

From our perspective, the latter approach seems to be the most promising, however, the optical injection-based encoding methods proposed in \cite{Yuan2016,Roberts2018,Paraiso2019,Lo2023simplified} place high demands on the electrical signal generator, amplifiers, and laser drivers, which must have a high bandwidth and work with analog signals of a rather complex shape. Adding decoy states \cite{Ma2005practicalDS} further complicates the optical circuit of the transmitter, which should include either an intensity modulator or an interferometer \cite{Lo2023simplified}. These features negate the advantages of this encoding method, increase the cost of the equipment, and complicate its practical use.

In this paper, we consider a method of time-bin encoding using pulsed optical injection, which ensures the preparation of quantum states without analog signals of complex shape. At the same time, since typical distances between nodes of MAN range from 5 to 20 km, it seems reasonable to abandon the decoy states, which, together with the proposed method of quantum state preparation, allows making the transmitter extremely simple.

Section \ref{sec:generalDescription} provides a general description of the proposed method. In section \ref{sec:simulations}, we show the results of the simulation, while the results of an experiment confirming the operating principle are reported in section \ref{sec:experiment}. Finally, in section \ref{securityAnalysis}, we analyze the security of the decoy-free three-state QKD protocol that can be implemented with the proposed method.

\section{Setup description}\label{sec:generalDescription}
\begin{figure}[t]
	\includegraphics[width=\columnwidth]{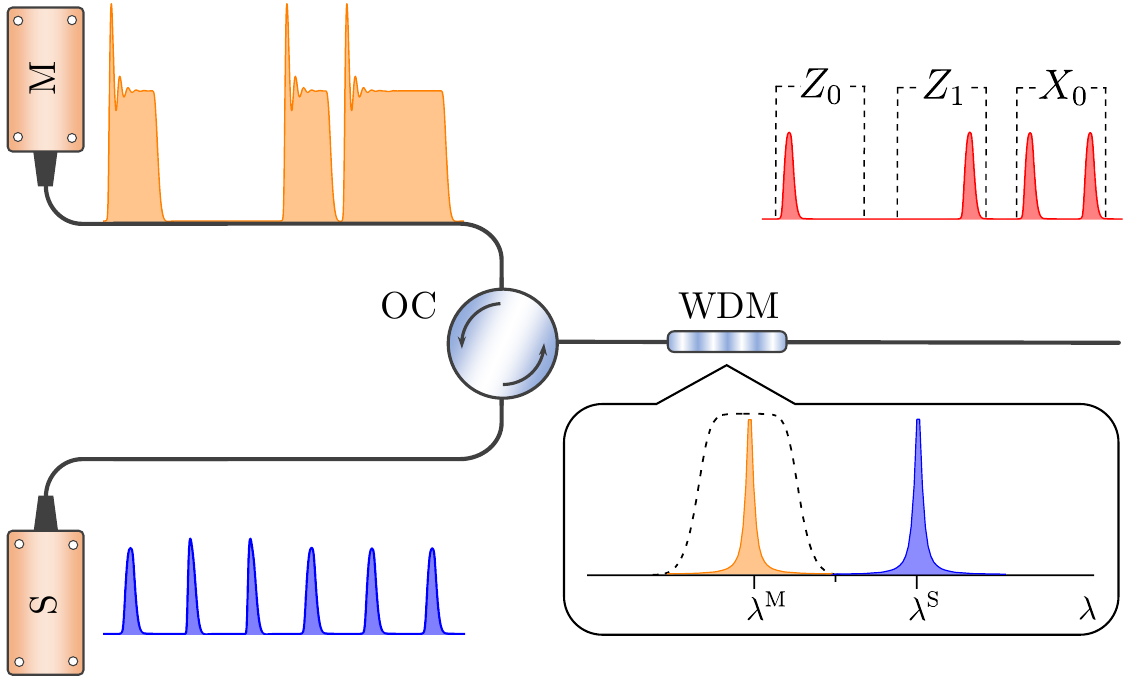}
	\caption{\label{fig:generalScheme} A simplified fiber-optic schematic of the transmitter without electro-optic modulators. M and S denote the master and slave lasers, respectively, OC is the optical	circulator, and WDM is the optical WDM filter. Optical pulses are shown next to the corresponding laser diodes. A schematic representation of the laser spectra is shown below the WDM filter (the filter passband is shown by the dotted line); $\lambda^\text{M}$ and $\lambda^\text{S}$ are the wavelengths of the master and slave lasers. The resulting optical signal after spectral filtering is shown above the spectra (the encoded states are highlighted by the dotted rectangles).}
\end{figure}

A simplified schematic of a fiber-optic transmitter without electro-optic modulators, implementing time-bin encoding, is shown in Fig.~\ref{fig:generalScheme}. The master and slave lasers are connected via an optical circulator, which third output is connected to an optical filter (WDM filter). (It is assumed that the fiber-optic outputs of the lasers and the circulator are made of the polarization maintaining fiber.) The wavelengths of the master and slave lasers are spectrally separated so that they fall into different WDM channels, and the filter passband is selected such that the radiation of the master passes through the filter, and the radiation of the slave laser is blocked.

The slave laser operates in a gain-switched mode and generates a regular sequence of short pulses with a repetition rate of $f_p$. The master laser also operates in a gain-switched mode and emits two types of pulses: 1) short pulses, with the duration approximately equal to the pulse repetition period of the slave laser, and 2) long pulses, with the duration to be approximately twice the pulse repetition period of the slave laser. The short pulses of the master are used to prepare states in the $Z$-basis, whereas long pulses are used to prepare states in the $X$-basis.

Recall that with the time-bin encoding the values of the bits in the $Z$-basis are specified by the time of the pulse appearance: \textquoteleft 0\textquoteright\ can be assigned to the state when the pulse appears in the early time bin ($Z_0$-state), and \textquoteleft 1\textquoteright\ can be assigned to the state when the pulse appears in the late time bin ($Z_1$-state) (see Figs.\,\,\ref{fig:generalScheme}, \ref{fig:simulationResults}, \ref{fig:experimentalResults}). In the $X$-basis, pulses appear in both time bins, and the bits are encoded by the phase difference between the pulses.

The encoding in the circuit schematically shown in Fig.\,\,\ref{fig:generalScheme} occurs as follows. When the master generates an optical pulse, the radiation in the corresponding pulse of the slave laser changes its wavelength, adjusting to the wavelength of the master laser due to the locking effect \cite{Shakhovoybook} --- such a pulse passes through the optical filter. If the pulse of the slave laser appears in the absence of optical injection, it will be blocked by the filter. Thus, by generating short pulses by the master at the right moments in time, one can create a sequence of bits in the $Z$-basis. 

To encode bits in the $X$-basis, the master laser must generate long pulses that simultaneously capture a pair of pulses of the slave laser: in this case, both pulses of the slave laser will pass through the optical filter. As explained in \cite{Shakhovoy2021direct}, the phase difference between a pair of adjacent slave laser pulses in this case is determined by the phase evolution of the electric field in the master laser pulse. The field phase, in turn, depends on the pump current, which can be used for encoding. In the proposed scheme, however, it is assumed that only one state, $X_0$, is prepared in the $X$-basis, so there is no need to change the pump current for long pulses, which allows digital signals to be used to modulate both lasers. 

The prepared laser pulses are attenuated to a quasi-single-photon level and sent to the quantum channel.

\section{Simulation}\label{sec:simulations}

\begin{table}[b]
	\caption{Lasers' parameters used for simulations.}
	\begin{center}
		\begin{tabular}{lc}
			\hline \hline \rule{0mm}{3mm}
			\textrm{Parameter}&
			\textrm{Value}\\
			\hline
			\rule{0mm}{3mm}
			Photon lifetime $\tau_\text{ph}$, ps & 1.0 \\
			\rule{0mm}{3mm}
			Electron lifetime $\tau_e$, ns & 1.0\\
			\rule{0mm}{3mm}
			Quantum differential output $\eta$& 0.3 \\
			\rule{0mm}{3mm}
			Threshold carrier number $N_\text{th}$
			&$\text{4.0}\times10^7$\\ \rule{0mm}{3mm}
			Transparency carrier number $N_\text{tr}$ &
			$\text{5.5}\times10^7$\\ \rule{0mm}{3mm}
			Photon energy $\hbar\omega_0$, eV & 0.8 \\
			\rule{0mm}{3mm}
			Spontaneous emission coupling factor $C_\text{sp}$ &
			$10^{-5}$\\ \rule{0mm}{3mm}
			Confinement factor $\Gamma$  & 0.12\\
			\rule{0mm}{3mm}
			Linewidth enhancement factor $\alpha$ & 5\\
			\rule{0mm}{3mm}
			Master gain compression factor $\gamma_P^\text{M}$, W$^{-1}$ &
			30\\ \rule{0mm}{3mm}
			Slave gain compression factor $\gamma_P$, W$^{-1}$ & 20\\
			\rule{0mm}{3mm}
			Master-slave coupling factor $\kappa_\text{inj}$, GHz & 200 \\
			\rule{0mm}{3mm}
			Master-slave detuning $\Delta \omega / 2\pi$, GHz & $-100$ \\
			\hline \hline
		\end{tabular}\label{tab:laserParameters}
	\end{center}
\end{table}

To demonstrate the proposed encoding method, we first performed numerical simulation of the \enquote{master\,+\,slave} laser system shown in Fig.~\ref{fig:generalScheme}. For this, we used the standard model for semiconductor lasers with optical injection
\cite{Shakhovoy2021direct, Shakhovoybook}, namely the system of
differential equations for the master:
\begin{eqnarray}
	\frac{dN^\text{M}}{dt} &=& \frac{I}{e} - \frac{N^\text{M}}{\tau_e^\text{M}} - \frac{Q^\text{M}G^\text{M}}{\Gamma^\text{M} \tau_\text{ph}^\text{M}}, \label{eq:NMaster} \\
	\frac{Q^\text{M}}{dt} &=&  (G^\text{M} - 1)\frac{Q^\text{M}}{\tau_\text{ph}^\text{M}} +	C_\text{sp}^\text{M}\frac{N^\text{M}}{\tau_e^\text{M}},\\
	\frac{\varphi^\text{M}}{dt} &=&  \frac{\alpha^\text{M}}{2 \tau_\text{ph}^\text{M}} (G_L^\text{M}-1),
\end{eqnarray}
and the corresponding system for the slave:
\begin{eqnarray}
	\frac{dN}{dt} &=& \frac{I}{e} - \frac{N}{\tau_e} - \frac{Q G}{\Gamma \tau_\text{ph}},\\
	\frac{dQ}{dt} &=& (G-1)\frac{Q}{\tau_\text{ph}} + C_\text{sp}\frac{N}{\tau_e}+ \nonumber\\
	&+& 2\kappa_\text{inj}\sqrt{Q^\text{M}Q}\cos(\Delta \omega t + \varphi^\text{M} - \varphi),  \\
	\frac{d\varphi}{dt}& =& \frac{\alpha}{2 \tau_\text{ph}}(G_L-1) + \nonumber\\
	&+& \kappa_\text{inj} \sqrt{\frac{Q^\text{M}}{Q}} \sin{(\Delta \omega t + \varphi^\text{M} - \varphi) } \label{eq:PhiSlave}
\end{eqnarray}
where the superscript M means <<master>>. In the equations \eqref{eq:NMaster}--\eqref{eq:PhiSlave}, $N$ is the number of carriers in the active layer of the laser, $Q$ is the normalized intensity of the electromagnetic field in the resonator, corresponding to the average photon number, $\varphi$ is the phase of the field, $I$ is the pump current, $e$ is the electron charge, $G_L$ is the linear dimensionless gain, defined by the relation
\begin{equation*}
	G_L=\frac{N - N_\text{tr}}{N_\text{th}-N_\text{tr}},
\end{equation*}
where $N_\text{th}$ is the number of carriers at threshold, and $N_\text{tr}$ is the number of carriers at transparency. The gain nonlinearity was taken into account using the formula ${G=G_L/\sqrt{1+2\gamma_P P}}$, where ${P=Q(\eta\hbar\omega_0/2\Gamma\tau_\text{ph})}$ is the measured optical power of the laser (the coefficient $1/2$ accounts for the fact that the power exits through both facets, but is generally measured only through one), and $\gamma_P$ is the gain compression factor. Other parameters: $\hbar\omega_0$ --- photon energy, $\eta$ --- quantum differential output, $\Gamma$ --- confinement factor, $C_\text{sp}$ --- spontaneous emission coupling factor (fraction of spontaneously emitted photons coupled into the lasing mode), $\alpha$ --- linewidth enhancement factor (Henry's factor), $\tau_\text{ph}$ --- photon lifetime, $\tau_e$ --- carrier lifetime, $\Delta\omega$ --- lasers' detuning, $\kappa_\text{inj}$ -- master-slave injection coupling factor.

In the simulation, the pump current was specified as a train of rectangular pulses; the current parameters were chosen to ensure gain switching for both lasers and stable frequency locking. The laser parameters used for the simulation are listed in Table\,\,\ref{tab:laserParameters} (the same parameters were used for both master and slave lasers, except for the gain compression factor.)

\begin{figure}[t]
	\includegraphics[width=\columnwidth]{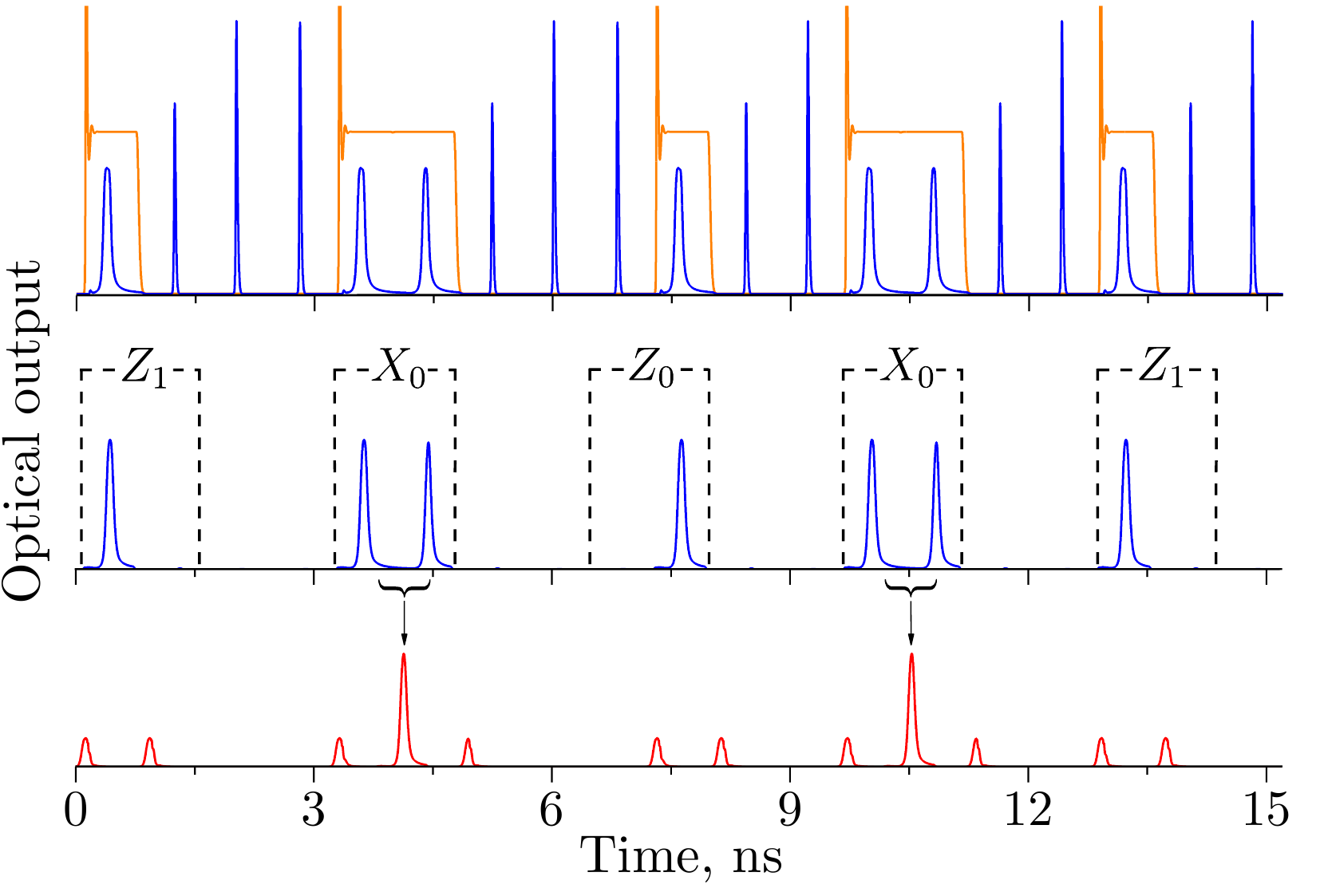}
	\caption{Numerical simulation demonstrating the proposed encoding method. The slave and master laser pulses are shown at the top. The optical filtering result is shown in the middle. The interference result is shown at the bottom.\label{fig:simulationResults}}
\end{figure}

The simulation results are shown in Fig.\,\,\ref{fig:simulationResults}. It was assumed that Alice prepares the following sequence of states: $Z_0$, $X_0$, $Z_1$, $X_0$, $Z_0$. For convenience, an additional delay of $2T$ was used between adjacent states, where $T=1/f_p$ is the pulse repetition period of the slave laser (the latter was set to 800\,ps, which corresponds to $f_p=1.25$\,GHz). Note that in general this delay is excessive, and in simulations it was used purely for demonstration purposes, since it helps to obtain a more visual result during decoding. (From an experimental point of view, however, such a delay may be useful to reduce the so-called intersymbol interference effect, which we discuss briefly in section \ref{sec:experiment}.)

Figure\,\,\ref{fig:simulationResults} (top) shows the pulse trains of the slave and master lasers. One can observe that when the slave laser emits pulses without optical injection, the relaxation spike exhibits a higher intensity compared to the case with master laser radiation. This is a consequence of the partial suppression of transients due to optical injection \cite{Shakhovoybook}. The middle part of Fig.\,\,\ref{fig:simulationResults} shows the result of optical filtering, simulated using a second-order Butterworth filter. As seen in the figure, the filter transmits only pulses for which frequency locking is achieved, yielding the desired sequence of states (the time windows corresponding to the prepared states are highlighted by dashed rectangles). Finally, the bottom panel of Fig.\,\,\ref{fig:simulationResults} presents the calculated interference of the generated pulse sequence with itself in an unbalanced interferometer with a delay line equal to $T$.

The next section presents experimental results that, as will be shown, agree well with the model calculations.

\section{Experiment}\label{sec:experiment}

\begin{figure}[t]
	\includegraphics[width=\columnwidth]{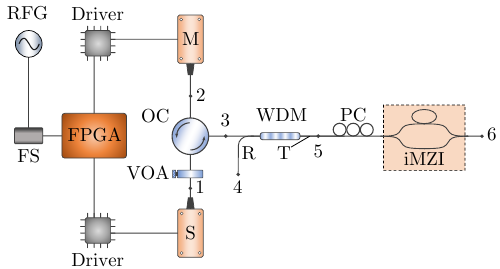}
	\caption{\label{fig:experimentalSetup} Schematic of an experimental setup: VOA --- variable optical attenuator, PC --- polarization controller, iMZI --- integrated Mach-Zehnder interferometer, FPGA --- field programmable gate array, FS --- frequency synthesizer, RFG --- reference frequency generator. R and T denote the reflection and transmission channels of the WDM filter, respectively. Other designations are as in Fig.~\ref{fig:generalScheme}.}
\end{figure}

The schematic of an experimental setup is shown in Fig.\,\,\ref{fig:experimentalSetup}. Two distributed feedback laser diodes optically coupled through a circulator were used for the experiment. A Shengshi DFB laser diode with a wavelength of 1550~nm in a 14-pin butterfly package with a built-in optical isolator was used as a master (M), and an Agilecom DFB laser diode without a built-in optical isolator was used as a slave laser (S). A variable optical attenuator (VOA) was installed between the master and slave lasers to control the injected optical power. A standard WDM filter with a bandwidth of 100~GHz and a central wavelength of 1549.32~nm (C35) was installed at the output of the circulator. The filtered output (channel T in Fig.\,\,\ref{fig:experimentalSetup}) passed through a polarization controller (PC) to set the required state of polarization for input into the integrated interferometer. A variable unbalanced integrated Mach-Zehnder interferometer thermally stabilized via a Peltier element (see \cite{Shakhovoy2023} for details) was used to observe pulse interference. To detect optical signals, a Thorlabs PDA8GS photodetector was used, and to obtain optical spectra, a Finisar WaveAnalyser 200A optical spectrum analyzer was used, which were installed at different points of the circuit, numbered in Fig.~\ref{fig:experimentalSetup}.

To modulate the pump current on both the master and slave lasers, pulsed laser drivers from \enquote{QRate} were used, based on the standard Texas Instruments ONET1151L chip. The control pulses were generated using high-speed (10\,Gbps) transceivers of the field-programmable gate array (FPGA), using a 156.25\,MHz clock signal from a frequency synthesizer (FS) from Silicon Labs, which, in turn, used the output of a high-stability oscillator (RFG) at a frequency of 10 MHz as a reference.

\begin{figure}[t]
	\includegraphics[width=\columnwidth]{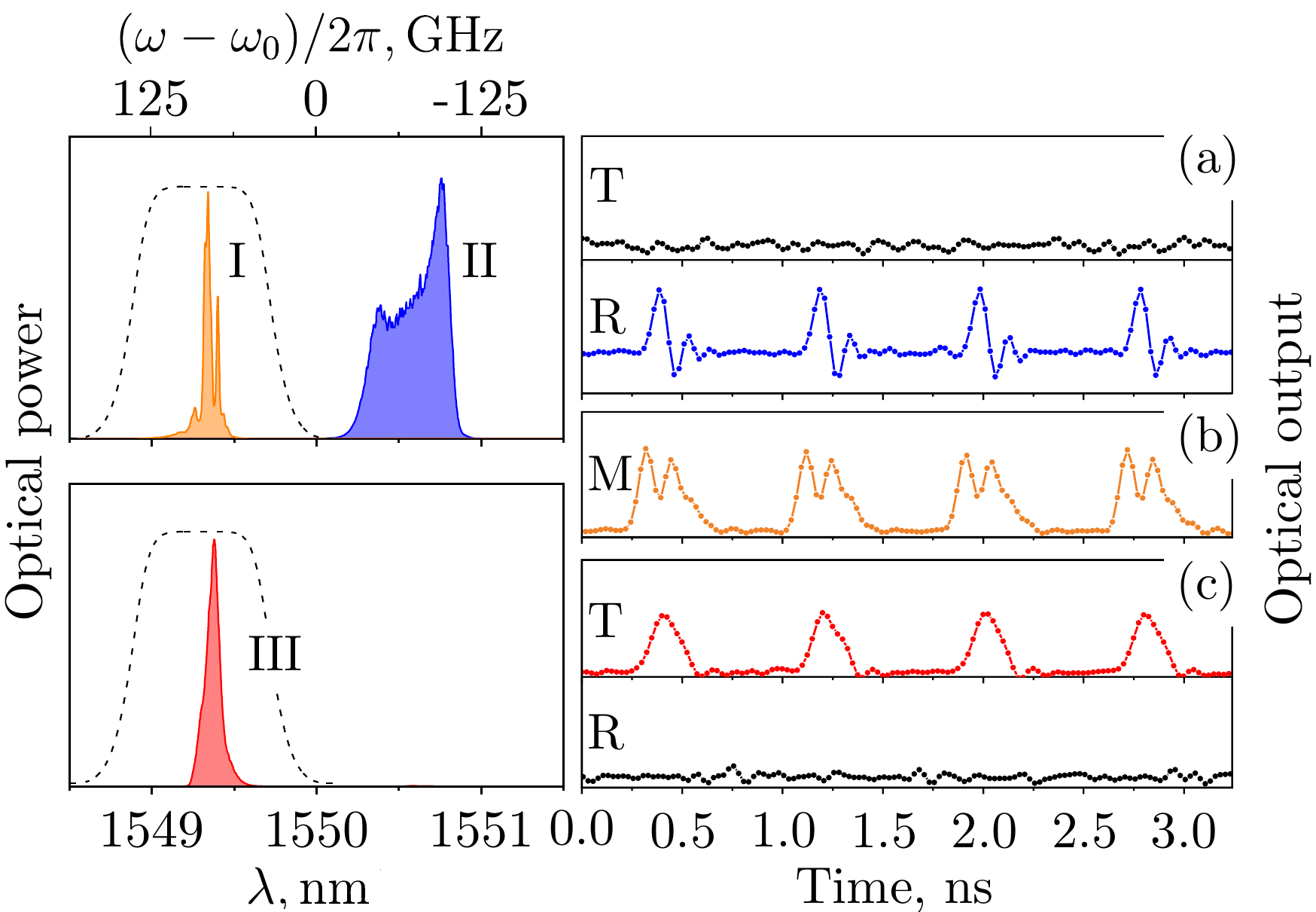}
	\caption{\label{fig:spectraAndPulses} Optical spectra and pulses of the master and slave lasers: (a) slave laser pulses when the master is switched off; (b) master laser pulses; (c) slave laser pulses under pulsed optical injection. I --- spectrum of a signal (b); II --- spectrum of a signal (a); III --- spectrum of a signal (c). R and T denote the reflection and transmission channels of the WDM filter, respectively; M denotes the master's output signal.}
\end{figure}

First, we recorded the spectra and optical signals of the master and slave lasers at a pulse repetition frequency of 1.25\,GHz. The output signal of the slave laser (with the master switched off) in both the reflection (R) and transmission (T) channels of the WDM filter (points 4 and 5, respectively, in Fig.~\ref{fig:experimentalSetup}), is shown in Fig.~\ref{fig:spectraAndPulses}(a). The corresponding spectrum of the slave laser (spectrum II) is displayed to the left of the pulses in Fig.~\ref{fig:spectraAndPulses}. As evident from the figure, the slave laser generates short optical pulses and exhibits a broad spectrum that falls outside the transmission window of the optical filter (indicated by the dotted line). The spectral width and shape suggest significant chirp, which is typical for short pulses of a gain-switched laser.

The master laser pulses, recorded at point 2 in Fig.~\ref{fig:experimentalSetup}, are shown in Fig.~\ref{fig:spectraAndPulses}(b); spectrum I corresponds to this signal. The master laser exhibits a narrower spectrum, though it contains distinct peaks associated with relaxation spikes in the pulse waveform. Notably, while both slave (Fig.~\ref{fig:spectraAndPulses}(a)) and master (Fig.~\ref{fig:spectraAndPulses}(b)) lasers were driven by identical 400\,ps rectangular electrical pulses, the master laser produced significantly longer optical pulses due to its higher bias current.

The signal of the slave laser subject to pulsed optical injection is shown in Fig.~\ref{fig:spectraAndPulses}(c). It was again measured at points 4 and 5 of the experimental setup (Fig.~\ref{fig:experimentalSetup}), but since the wavelength of the optical signal was now locked to the master, it was not blocked by the filter. In addition, the pulses of the slave laser became longer due to partial suppression of the relaxation spike. The spectrum became significantly narrower (see spectrum III in Fig.~\ref{fig:spectraAndPulses}) indicating a reduction in chirp. These results confirm successful frequency locking under pulsed optical injection, enabling time-bin encoding implementations.

\begin{figure}[t]
	\includegraphics[width=\columnwidth]{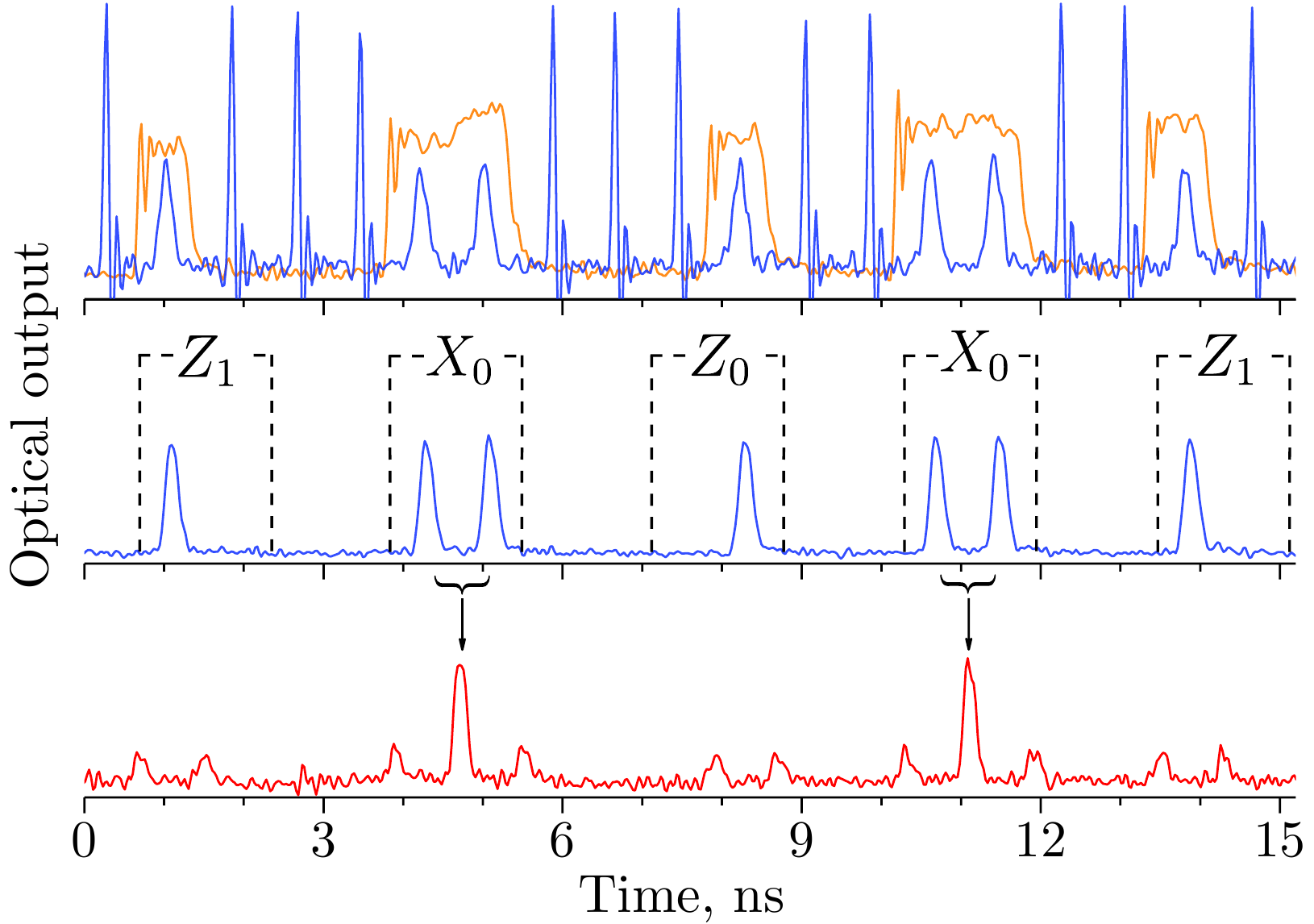}
	\caption{\label{fig:experimentalResults}Experimental results demonstrating the proposed encoding method. The slave and master laser pulses are shown at the top. The result of optical filtering is shown in the middle. The interference result is shown at the bottom.}
\end{figure}

An experimental demonstration of the proposed encoding method is shown in Fig.~\ref{fig:experimentalResults}. We implemented a short pseudo-random sequence repeated every 8 states (a 5-state subsequence matching the simulation in Fig.~\ref{fig:simulationResults} is shown). The slave laser operated at 1.25\,GHz pulse repetition rate, while the state preparation rate was 312.5\,MHz due to an additional 1.6\,ns inter-state delay. The top of Fig.~\ref{fig:experimentalResults} shows the signals from the master and slave lasers, recorded at points 2 and 3 of the experimental setup, respectively (see Fig.~\ref{fig:experimentalSetup}). One can see that in the presence of the master radiation, the pulses of the slave laser become much wider and their amplitude is significantly reduced. In the middle of Fig.~\ref{fig:experimentalSetup}, the slave laser signal after the WDM filter is shown, recorded at point 5 of the schematic. It is evident that the slave laser pulses generated in the absence of the master's radiation are filtered quite effectively.

To verify correct phase preparation in the $X$-basis, we employed an unbalanced Mach-Zehnder interferometer with an 800\,ps delay line. The interferometer’s additional phase shift was controlled via temperature adjustment of the photonic integrated circuit, calibrated to produce constructive interference for $X_0$-state pulse pairs. The result of the interference is shown in Fig.~\ref{fig:experimentalResults} at the bottom. All prepared $X_0$ states in the sequence (only two of them are shown in Fig.~\ref{fig:experimentalResults}) corresponded to constructive interference, which indicates the reliability of the encoding.

It is important to note that the shape of the master pulses can vary significantly depending on their preceding state(s), an effect known in the literature as intersymbol interference. In particular, at the state preparation frequency of 625\,MHz, we observed significant intersymbol interference that distorted the master pulse shapes and did not allow us to stably prepare states in the $X$-basis. In order to minimize the dependence of the signal on the \enquote{prehistory}, we introduced the additional delay between states. Although in the presence of such a delay the master pulse shape was still noticeably distorted (see the master signal in Fig.~\ref{fig:experimentalResults}), the influence of uncontrolled transients was significantly reduced, and we were able to select modulation current parameters that provide stable coding.

Two primary factors may contribute to the observed intersymbol interference. The first of them is of a purely physical nature and is associated with the finiteness of the carrier lifetime \cite{Petermannbook}. In order to level it out, it is necessary to set the bias current as close as possible to the threshold, so that the number of carriers $N$ does not have time to decrease significantly between the pump current pulses. Here, however, one should be careful, since high values of the bias current can lead to phase correlations between the pulses \cite{Shakhovoy2023}, which, in turn, can negatively affect the security of the QKD \cite{Kobayashi2014}. In our case, due to the relatively low repetition rate of the master pulses (312.5\,MHz), the influence of the finite carrier lifetime can be neglected.

Another reason is purely technical and is caused by inaccuracies in the design of the laser driver, which lead to impedance mismatch in different sections of the electrical circuit. We believe that in our case, this is the main cause of intersymbol interference. Note, however, that no intersymbol interference was observed in the slave laser signal without master, since a regular pulse train was used to pump it. Moreover, the distortion of the master pulse shape has practically no effect on the intensity and width of the slave laser pulses, in other words, the intersymbol interference in the master signal does not lead to any noticeable consequences in the $Z$-basis, which guarantees a low level of errors in the quantum key even when using low-quality signals. This feature is an important advantage of the coding method we propose.

\section{Security analysis}\label{securityAnalysis}
As mentioned in previous sections, the transmitter under consideration allows Alice to generate only three different states: two in the $Z$-basis and one in the $X$-basis. This restriction prevents the implementation of a standard four-state BB84 protocol \cite{BB84, GLLP}; therefore, we will focus on analyzing a \textit{three-state BB84-type protocol} \cite{fung2006security}.

Note that by a \enquote{QKD protocol} we mean here a specific implementation of a key distribution method, including a specified error correction procedure, a method for preparing and measuring quantum states, privacy amplification (a certain formula for the secret key rate), etc. Therefore, for greater generality, we will talk below about \textit{families} of protocols. So, we will briefly analyze here the security of the family of three-state BB84-type QKD protocols that can be implemented using the proposed transmitter.

\subsection{Protocol formalization}
We will consider the family of QKD protocols, formalized by the following set of basic steps:

\begin{protocol}
	\begin{enumerate}[leftmargin=*]
		\item Alice generates a random string $\vec{s}$ of length $N$, where ${s_i \in \mathfrak{S} = \left\{0, 1, +\right\}}$.  The characters \textquoteleft 0\textquoteright\ and \textquoteleft 1\textquoteright\ are chosen with the probability $p^A_Z/2$, and the character \textquoteleft +\textquoteright\ is chosen with the probability $p^A_X$. Here, $p^A_Z$ corresponds to the Alice’s probability of choosing the $Z$-basis, and $p^A_X$ is the probability of choosing the $X$-basis. Based on $\vec{s}$, Alice prepares $N$ quantum states $\ket{\psi_{s_i}}$ and sends them to Bob via the quantum channel. 
		
		\item Bob generates a random string $\vec{b}$ of length $N$, where a character ${b_i \in \{Z,X\}}$ is chosen with probability $p^B_Z$ for $Z$ or $p^B_X$ for $X$. According to $\vec{b}$ Bob selects measurement bases for the incoming states. (When measuring in the $X$-basis, the outcome will be either the state $\ket{\psi_+}$ sent by Alice or an orthogonal state $\ket{\psi_-}$, which Alice does not send.)
		
		\item Alice and Bob publicly compare the chosen bases. To do this, Alice announces the result $\vec{a}$ of a mapping
		\begin{equation*}
			a_i =\begin{cases} 
				Z, & s_i \in \{0,1\}, \\ 
				X, & s_i = +,
				\end{cases}
		\end{equation*}
		and Bob reveals $\vec{b}$. All events satisfying ${a_i \neq b_i}$, for which the bases do not match, are discarded. For events that satisfy ${a_i = b_i=Z}$, Alice writes the corresponding value $s_i\in\{0,1\}$ into the bit string	$\vec{\alpha}'$, and Bob writes the measurement result into the bit string $\vec{\beta}'$. In the case of ${a_i=b_i=X}$, Bob stores the measurement result in the bit string $\vec{\beta}''$.
		
		\item Alice and Bob evaluate the error rate in their sifted keys $\vec{\alpha}'$ and $\vec{\beta}'$, and then perform error correction. As a result, they obtain identical bit strings	${\vec{\alpha} = \vec{\beta}}$ of length $M \leq N$ with high probability.
		
		\item Bob publicly announces $\beta_i''$. The error rate in the $X$-basis is determined.
		
		\item Alice and Bob perform privacy amplification on the sifted and corrected keys $\vec{\alpha}=\vec{\beta}$ and obtain an identical secret key $\vec{r}$.
	\end{enumerate}
\end{protocol}

When using time-bin encoding, Alice prepares three states ($\ket{\psi_{0}}$, $\ket{\psi_{1}}$, and $\ket{\psi_+}$) \enquote{living} in the extended Hilbert space of two temporal modes, which we will call the \textit{early} and \textit{late} modes, or the $Z_0$- and $Z_1$-modes, respectively. Each state can be then written as a tensor product:
\begin{equation}\label{eq:psiStates}
	\begin{split}
		\ket{\psi_0}& \equiv \iket{\sqrt{\mu} e^{i\varphi}}{Z_0} \otimes
		\iket{0}{Z_1}, \\
		\ket{\psi_1}& \equiv \iket{0}{Z_0} \otimes \iket{\sqrt{\mu}
			e^{i\varphi}}{Z_1}, \\
		\ket{\psi_+}& \equiv \iket{\sqrt{\nu/2}\, e^{i\varphi}}{Z_0}
		\otimes \iket{\sqrt{\nu/2}\, e^{i\varphi}}{Z_1},
	\end{split}
\end{equation}
where $\iket{\sqrt{\mu}e^{i\varphi}}{Z_0/Z_1}$ is a coherent state in an early/late temporal mode and $\iket{0}{Z_0/Z_1}$ is a vacuum state in the corresponding mode. Here, $\sqrt{\mu}$ represents the amplitude of the coherent state in the $Z$-basis ($\mu$ is the intensity corresponding to the average photon number), and $\varphi$ denotes the phase. For the sake of generality, we assume that the intensity of the coherent states in early and late modes of the $X$-basis may differ from that in the $Z$-basis, i.\,e., in general case, ${\mu\ne\nu/2}$. Since the lasers in the transmitter under consideration operate under gain-switching, we assume $\varphi$ is a uniformly distributed random variable, thus satisfying the phase randomization condition \cite{Shakhovoy2023}.

Next, we expand the coherent states in the Fock basis and average them over the phase $\varphi$. For the state $\ket{\psi_+}$ in Eq.\,\eqref{eq:psiStates}, the resulting phase-averaged state would be quite cumbersome. Therefore, we modify our formalism by introducing creation operators for states in the $X$-basis:
\begin{equation}
	\begin{split}
		a^\dagger_{X_0} &\equiv \frac{1}{\sqrt{2}}\left( a^\dagger_{Z_0}
		+ a^\dagger_{Z_1} \right), \\
		a^\dagger_{X_1} &\equiv \frac{1}{\sqrt{2}}\left( a^\dagger_{Z_0}
		- a^\dagger_{Z_1} \right).
	\end{split}
\end{equation}
These operators correspond to two orthogonal modes and induce a \enquote{rotated} Fock basis:
\begin{equation}
	\begin{gathered}
		\iket{n}{X_0}\otimes \iket{m}{X_1} \equiv
		\frac{(a^\dagger_{X_0})^n}{\sqrt{n!}}
		\frac{(a^\dagger_{X_1})^m}{\sqrt{m!}} \ket{\Omega},
	\end{gathered}
\end{equation}
where $\ket{\Omega}\equiv \ket{0}_{Z_0}\otimes \ket{0}_{Z_1}$ is the vacuum state. The average states emitted by Alice form a mixture of $k$-photon states with Poissonian statistics:
\begin{equation}
	\label{eq:poisson_states}
	\begin{gathered}
		\rho_0 = \sum\limits_{k=0}^\infty e^{-\mu} \frac{\mu^k}{k!}
		\kbri{k}{k}{Z_0} \otimes \kbri{0}{0}{Z_1}, \\
		\rho_1 = \sum\limits_{k=0}^\infty e^{-\mu} \frac{\mu^k}{k!}
		\kbri{0}{0}{Z_0} \otimes \kbri{k}{k}{Z_1}, \\
		\rho_+ = \sum\limits_{k=0}^\infty e^{-\nu} \frac{\nu^k}{k!}
		\kbri{k}{k}{X_0} \otimes \kbri{0}{0}{X_1}.
	\end{gathered}
\end{equation}

To prove the security of the family of protocols under consideration, we must estimate the correlation between the string $\vec{\alpha}$ and an eavesdropper's quantum system. However, performing such analysis for states \eqref{eq:poisson_states} --- which are defined in an infinite-dimensional Hilbert space --- is generally intractable. We therefore reduce this problem to the well-established security proof for a family of protocols that use states in a finite-dimensional Hilbert space.

Let us denote the space of two modes $Z_0$ and $Z_1$ as $\mathcal{H}_{A'}$ and introduce a classical register \enquote{living} in Hilbert space $\mathcal{H}_A$ with an orthonormal basis $\{\iket{j}{A}\}_{j\in \mathfrak{S}}$, where $\mathfrak{S} = \{0,1,+\}$. Only states prepared and measured in the same basis contribute to the sifted key (we call these events successful); the probabilities of such events are given by
\begin{equation}\label{eq:probabilities}
	\begin{aligned}
		p_0 = p_1  = \frac{p_Z}{2} &= \frac{1}{2} \frac{p^A_Z
			p^B_Z}{p^A_Z p^B_Z + p^A_X p^B_X}, \\
		p_+ = p_X &= \frac{p^A_X p^B_X}{p^A_Z p^B_Z + p^A_X p^B_X}.
	\end{aligned}
\end{equation}
Since we consider only successful events, the preparation of states defined in Eq.\,\eqref{eq:poisson_states} can be interpreted as a measurement performed on the classical subsystem $A$ of the quantum-classical system $AA'$, characterized by the density operator
\begin{equation}
	\rho^{AA'} \equiv \sum_{j\in \mathfrak{S}} p_j \kbri{j}{j}{A}
	\otimes \rho^{A^{\prime}}_{j},
\end{equation}
where $\rho^{A^{\prime}}_j$ are defined by Eq.\,\eqref{eq:poisson_states}, and $p_j$ are probabilities given by Eq.\,\eqref{eq:probabilities}.

Let $\dmspace{\mathcal{H}}$ be a space of all density operators over $\mathcal{H}$. Consider some projective measurement ${\Pi:\dmspace{\mathcal{H}_{A^{\prime}}} \rightarrow \dmspace{\mathcal{H}_{A^{\prime}}}}$, described by a set of two projectors $\{\mathcal{P}_\text{sec}, \mathcal{P}_\text{non}\}$ with outcomes \enquote{sec} and \enquote{non}, respectively, such that
\begin{equation}\label{eq:projectiveMeasurement}
	\bigl( \mathds{1}_A\otimes \Pi \bigr) \bigl(\rho^{AA'}\bigr) = \rho^{AA'}.
\end{equation}
Relation \eqref{eq:projectiveMeasurement} implies that Alice can perform measurement $\Pi$ on her states without causing distortion. For instance, quantum mechanics does not prohibit measuring the number of photons in coherent pulses and only then sending them to Bob, simultaneously informing him of the measurement results. Obviously, in this case, Alice's states will still be described (on average) by Eq.\,\eqref{eq:poisson_states}.

Let ${\mathcal{P}_\text{sec}: \mathcal{H}_{A^{\prime}} \rightarrow \mathcal{H}_\text{sec}}$ be a projector such that $\mathcal{H}_\text{sec}$ is a finite Hilbert space, and a security proof for some QKD protocol using quantum states from such a space is known. If Alice, after performing the measurement $\Pi$ and then sending her quantum states, tells Bob the state number, for which she obtained the outcome \enquote{sec} (we will call the corresponding states \emph{secret}), then they can simply discard all other sent states as \textit{non-secret}. As a result, they will have a smaller number of states that are characterized by the density operator
\begin{multline}\label{eq:tilderho}
	\!\!\tilde{\rho}^{AA'}\!\! = \!\frac{1}{\tr \left\{ \left(\mathds{1}_A\!\otimes\! \mathcal{P}_\text{sec}\right) \!\rho^{AA'} \right\}}	\!\sum\limits_{j\in \mathfrak{S}} p_j \!\kbri{j}{j}{A} \!\otimes\!	\mathcal{P}_\text{sec} \rho_j^{A'} \mathcal{P}_\text{sec} \\
	\equiv \sum\limits_{j\in \mathfrak{S}} \tilde{p}_j \kbri{j}{j}{A}\otimes \tilde{\rho}^{A'}_j.
\end{multline}
So, a family of QKD protocols that uses states from Eq.\,\eqref{eq:poisson_states} can always be formally reduced to a protocol in which Alice prepares the states
\begin{equation}\label{eq:BaseProtocol_states}
	\tilde{\rho}^{A'}_j \equiv \frac{\mathcal{P}_\text{sec} \rho^{A'}_j
		\mathcal{P}_\text{sec}}{\tr \{\mathcal{P}_\text{sec}\rho^{A'}_j\}},
\end{equation}
with the probabilities
\begin{equation}
	\label{eq:BaseProtocol_probs}
	\tilde{p}_j \equiv p_j \frac{\tr \{\mathcal{P}_\text{sec}
		\rho^{A^{\prime}}_j\} }{\sum_{i \in \mathfrak{S}} p_i \tr
		\{\mathcal{P}_\text{sec} \rho^{A^{\prime}}_i\}}
\end{equation}
that \enquote{live} in a finite Hilbert space. As we shall see below, the choice of the projector $\mathcal{P}_\text{sec}$ specifies the family of protocols and also determines what statistics of the measurement results Alice and Bob must compute in order to estimate the fraction of secret information in the sifted key.

As can be readily seen, the introduced set of operators $\{\mathcal{P}_\text{sec}, \mathcal{P}_\text{non}\}$ formally describes Alice's preparation of coherent states with varying photon numbers. The operator $\mathcal{P}_\text{sec}$ may be considered as a projector onto Fock states with ${n<2}$ photons (vacuum and single-photon states), which are \enquote{secret} in the sense that Eve cannot perform the photon-number-splitting attack on them. The operator $\mathcal{P}_\text{non}$ then projects coherent states onto Fock states with ${n\ge2}$. Therefore, we can define these operators as follows:
\begin{equation}\label{eq:projectors}
	{\mathcal{P}_\text{sec} = \sum_{n\in \mathcal{N}} \mathcal{P}_n},\quad
	{\mathcal{P}_\text{non} = \sum_{n\notin \mathcal{N}} \mathcal{P}_n},
\end{equation}
where
\begin{equation}
	{\mathcal{P}_n \equiv \sum_{k=0}^{n} \kbri{n-k}{n-k}{Z_0} \otimes
		\kbri{k}{k}{Z_1}}
\end{equation}
are projectors onto the $n$-photon subspaces of two temporal modes, while $\mathcal{N}$ denotes a subset of non-negative integers (when considering only vacuum and single-photon states as secret, we have $\mathcal{N}=\{0,1\}$).

We will use projectors of the form \eqref{eq:projectors} to analyze protocols with decoy states, and then show how this approach can be used to compute the secret key rate  without decoy states.

\subsection{Secret key rate with decoy states}
As is well known, QKD protocols using weak coherent pulses rather than single photons become vulnerable to photon-number-splitting attack \cite{Huttner95,Lutkenhaus2002}. This vulnerability arises because laser pulses follow Poissonian photon statistics, meaning some coherent states may contain multiple photons. In principle, an eavesdropper could extract one photon from each multi-photon pulse ($n>1$), wait for basis reconciliation, and then measure the remaining photons without introducing sifted-key errors. By additionally blocking all single-photon pulses (masking this as channel loss), Eve could gain complete knowledge of the secret key.

As a countermeasure, Alice can decrease the laser pulse intensity to make multi-photon pulses ($n>1$) statistically insufficient for a successful attack. This approach significantly reduces both the key rate and the distance, so instead of using a very low intensity, Alice and Bob can try to estimate the fraction of single-photon pulses among all states sent by Alice, and then, assuming that these states are secure, make the eavesdropper's information about the key negligible during privacy amplification. This approach allows for higher key rates and increases the QKD distance.

To estimate the fraction of single-photon pulses (and the corresponding error rate), it has been proven effective to use so-called decoy states (DS) \cite{Ma2005practicalDS} --- additional laser pulses of varying intensities. Alice randomly sends decoy states (interleaved with signal states) through the quantum channel to Bob. Later, during the basis reconciliation stage, she communicates over the public channel which intensities she selected. In practice, three different pulse intensities are typically used in each basis. Here, we assume that in the $Z$-basis, Alice employs intensities $\mu_0$, $\mu_1$, and $\mu_2$, while in the $X$-basis, she uses $\nu_0$, $\nu_1$, and $\nu_2$ (where $\mu_0$ and $\nu_0$ correspond to the intensities of signal states). The quantities of interest (for the $Z$-basis) can then be estimated using the well-known formulas \cite{Ma2005practicalDS}:
\begin{widetext}
	\begin{align}\label{eq:DecoyStates}
		Q_1^Z\! &\geq\! Q_1^{Z,L}\!=\! \mu_0 e^{-\mu_0} Y_{1}^{Z,L},  \quad
		E_{1}^Z \leq E_{1}^{Z,U}\! = \!
		\frac{E_{\mu_1}Q_{\mu_1}e^{\mu_1}\! -
			\!E_{\mu_2}Q_{\mu_2}e^{\mu_2}}{ (\mu_1 - \mu_2) Y_{1}^{Z,L}}, \nonumber \\
		Y_{1}^{Z,L} &= \frac{\mu_0}{\mu_0\mu_1 - \mu_0\mu_2 - \mu_1^2 +
			\mu_2^2}\left( Q_{\mu_1}e^{\mu_1} - Q_{\mu_2} e^{\mu_2} -
		\frac{\mu_1^2-\mu_2^2}{\mu_0^2}\left(Q_{\mu_0}e^{\mu_0} -
		Y_{0}^{Z,L} \right) \right), \\
		Y_{0}^{Z,L} &= \max\left\{ \frac{\mu_1 Q_{\mu_2}e^{\mu_2}  -
			\mu_2 Q_{\mu_1}e^{\mu_1} }{\mu_1 - \mu_2}, 0 \right\}, \nonumber
	\end{align}
\end{widetext}
where $Q_\gamma$ is the experimentally determined gain (the fraction of registered states with intensity $\gamma$), $E_\gamma$ is the corresponding bit error rate (also measured experimentally), $Q_1^Z$ is the single-photon gain in the $Z$-basis, $E_1^Z$ is the corresponding single-photon bit error rate, and $Y_n^Z$ is the yield (the probability of the  detector's click given that an $n$-photon pulse was sent). The superscripts $U$ and $L$ denote the upper and lower bounds of the estimated quantity.

The formulas for the $X$-basis are obtained by replacing the superscript $Z$ with $X$ and the intensities $\mu_0$, $\mu_1$ and $\mu_2$ with $\nu_0$, $\nu_1$ and $\nu_2$, respectively, in \eqref{eq:DecoyStates}. Note that the estimates in \eqref{eq:DecoyStates} are valid under the assumptions ${0 \leq \mu_2 < \mu_1}$ and ${\mu_1+ \mu_2 < \mu_0}$ (and analogously for the $X$-basis intensities).

Using the projector
\begin{equation}
	\mathcal{P}_1 = \kbri{1}{1}{Z_0}\otimes \kbri{0}{0}{Z_1} +
	\kbri{0}{0}{Z_0}\otimes \kbri{1}{1}{Z_1}
\end{equation}
as $\mathcal{P}_\text{sec}$ in \eqref{eq:BaseProtocol_states} and \eqref{eq:BaseProtocol_probs}, we obtain the following states instead of \eqref{eq:poisson_states}:
\begin{equation}\label{eq:logicalstatesDS}
	\begin{split}
		\tilde{\rho}_{0}\! &= \!\kbri{1}{1}{Z_0}\! \otimes\!
		\kbri{0}{0}{Z_1}, \,\,\,\,
		\tilde{\rho}_{1}\!  =\! \kbri{0}{0}{Z_0}\! \otimes \!\kbri{1}{1}{Z_1}, \\
		\tilde{\rho}_{+}\! &=\! \frac{1}{2}\bigl(
		\kbri{1}{1}{Z_0}\!\otimes\!
		\kbri{0}{0}{Z_1}+\kbri{0}{0}{Z_0}\!\otimes\! \kbri{1}{1}{Z_1}+\\
		&+ \kbri{0}{1}{Z_0}\!\otimes\! \kbri{1}{0}{Z_1} +
		\kbri{1}{0}{Z_0}\!\otimes \!\kbri{0}{1}{Z_1}\bigr),
	\end{split}
\end{equation}
and corresponding probabilities:
\begin{equation}
	\label{eq:logicalprobabilitiesDS}
	\begin{aligned}
		\tilde{p}_{0}=\tilde{p}_{1} \propto \frac{p_Z}{2}
		e^{-\mu}\mu,\quad \tilde{p}_{+} \propto p_X e^{-\nu}\nu.
	\end{aligned}
\end{equation}

The security proof for the family of three-state protocols using states of the form \eqref{eq:logicalstatesDS} is well established  \cite{fung2006security} and provides the following formula for the secret key rate:
\begin{equation}\label{eq:keyRateDecoy}
	R = Q_{1}^{Z,L}\,r(E_{1}^{Z,U}, E_{1}^{X,U}) - f_\text{ec}
	Q_{\mu_0} h(E_{\mu_0}).
\end{equation}
Note that our state preparation probabilities \eqref{eq:logicalprobabilitiesDS} differ from 1/3 (the value used in \cite{fung2006security}); nevertheless, Eq.~\eqref{eq:keyRateDecoy} remains valid for the asymptotic secret key rate $R$ when $X$- and $Z$-bases are chosen with different probabilities, provided the states within the $Z$-basis are equiprobable. Since only one state is used in the $X$-basis, key bits can only be extracted from the $Z$-basis states (and only such bits require error correction).

In Eq.\,\,\eqref{eq:keyRateDecoy}, the first term represents the key reduction due do privacy amplification, while the second term accounts for the key leakage after error correction. The values of $Q_{1}^{Z,L}$, $E_{1}^{Z,U}$ and $E_{1}^{X,U}$ in \eqref{eq:keyRateDecoy} are determined through the decoy state method using relations \eqref{eq:DecoyStates}, with the gain $Q_{\mu_0}$ and the bit error rate $E_{\mu_0}$ in the $Z$-basis being measured experimentally. Here, $f_\text{ec}$ denotes the error correction efficiency coefficient  (typically ranging from 1.15 to 1.22), ${h(p)=-p\log(p)-(1-p)\log(1-p)}$ is the binary entropy, and the reduction factor $r$ is a function of two variables, ${r\equiv r(\omega,\theta)}$, defined by the relations \cite{fung2006security}:
\begin{equation}\label{eq:FLrate}
	r(\omega,\theta) = 1 - h(\kappa),\quad \kappa=\omega\cdot
	\max_{\delta\le 1}(\varepsilon^2+\delta^2),
\end{equation}
where
\begin{equation}\label{eq:phaseError}
	\begin{split}
		\varepsilon &=\theta\Bigl(
		\tilde{\theta}\delta+\sqrt{\tilde{\theta}(1-\delta^2)} +
		\Bigl[\tilde{\omega}(\tilde{\theta}+1)-  \\
		-&1-\delta^2(\tilde{\theta}-1)-2\delta\sqrt{\tilde{\theta}(1-\delta^2)}\Bigr]^{1/2}\Bigr),\\
		\tilde{\omega} &= \frac{1 - \omega}{\omega}, \quad \tilde{\theta}
		= \frac{1-\theta}{\theta},\quad\delta, \varepsilon\ge 0
	\end{split}
\end{equation}
(the parameter $\kappa$ represents the phase error rate, while $\omega$ and $\theta$ correspond to bit error rates in different bases).

\subsection{Secret key generation rate without decoy states}
The transmitter shown schematically in Fig.\,\,\ref{fig:generalScheme} cannot generate laser pulses with different intensities and thus cannot implement decoy-state protocols. Without decoy states, we cannot obtain reliable estimates for $Q_{1}^Z$, $E_{1}^Z$, and $E_1^X$. Consequently, we must consider the worst-case scenario, where all lost states were single-photon pulses, and all errors occurred only in single-photon states. Under these assumptions, we can only establish general inequalities (for brevity, we omit the basis superscript):
\begin{equation}
	\label{eq:woDS_estimations}
	\begin{split}
		Q_{1} &= Q_\gamma - Q_{0} - Q_{\geq 2}  = \\
		&= Q_{\gamma} - Y_{0} e^{-\gamma} - \sum\limits_{n=2}^\infty
		Y_{n} \frac{\gamma^n}{n!} e^{-\gamma}  \geq \\
		&\geq Q_{\gamma} - Y_{0} e^{-\gamma} - 1 + (1+\gamma)e^{-\gamma}, \\
		E_{1} Q_{1} &= E_{\gamma} Q_{\gamma} - \sum\limits_{n\neq 1}
		E_{n} Y_{n} \frac{\gamma^n}{n!} e^{-\gamma} \leq E_{\gamma} Q_{\gamma},
	\end{split}
\end{equation}
where we used the fact that ${0 \leq Y_{n} \leq 1}$ and ${0 \leq E_{n}}$. However, without decoy states, we cannot properly estimate the vacuum contribution ${Q_0=Y_0e^{-\gamma}}$ in the first inequality of Eq.\,\,\eqref{eq:woDS_estimations}. A more practical approach is to consider a lower bound on the combined gain ${Q_{0+1} \equiv 	Q_{0} + Q_{1}}$ instead of the single-photon gain $Q_1$. This bound can be obtained by moving $Q_0$ to the left-hand side of the inequality. Physically, this corresponds to treating both single-photon states and vacuum pulses (dark counts) as valid detection events. Formally, we implement this by setting the projection operator $\mathcal{P}_\text{sec}$ in \eqref{eq:tilderho} to $\mathcal{P}_0 + \mathcal{P}_1$.  In this case, Alice's states take the following form:
\begin{widetext}	
	\begin{equation}\label{eq:logicalstatesWODS}
		\begin{gathered}
			\tilde{\rho}_{0} \equiv\frac{1}{1 + \mu} \left[
			\vphantom{\sum}\kbri{0}{0}{Z_0} \otimes \kbri{0}{0}{Z_1} + \mu
			\kbri{1}{1}{Z_0} \otimes \kbri{0}{0}{Z_1} \right], \\
			\tilde{\rho}_{1} \equiv \frac{1}{1 + \mu} \left[
			\vphantom{\sum}\kbri{0}{0}{Z_0} \otimes \kbri{0}{0}{Z_1} + \mu
			\kbri{0}{0}{Z_0} \otimes \kbri{1}{1}{Z_1} \right], \\
			\tilde{\rho}_{+} \equiv \frac{1}{1 + \nu} \left[
			\kbri{0}{0}{Z_0} \otimes \kbri{0}{0}{Z_1} +
			\frac{\nu}{2}\Bigl(\iket{1}{Z_0}\iket{0}{Z_1}\! +
			\iket{0}{Z_0}\iket{1}{Z_1} \Bigr)\Bigl(
			\ibra{1}{Z_0}\ibra{0}{Z_1} + \!\ibra{0}{Z_0}\ibra{1}{Z_1} \Bigr)\right],
		\end{gathered}
	\end{equation}
\end{widetext}
and the probabilities are
\begin{equation}\label{eq:logicalprobabilitiesWODS}
	\tilde{p}_{0} = \tilde{p}_{1} \propto \frac{p_Z}{2} e^{-\mu}( 1 +
	\mu ),\quad \tilde{p}_{+} \propto p_X e^{-\nu}( 1 + \nu )
\end{equation}
(it is assumed that in the $Z$-basis coherent states have intensity $\mu$, while in the $X$-basis they have intensity $\nu$).

Using Eq.\,\,\eqref{eq:woDS_estimations}, we obtain the following estimates for the \enquote{zero\,+\,single-photon} gain and the corresponding bit error rate (in the $Z$-basis):
\begin{equation}
	\begin{gathered}
		Q^{Z,L}_{0+1} = Q_{\mu} - 1 + (1+\mu)e^{-\mu}, \\
		E_{0+1}^{Z,U}  = E_\mu Q_\mu/Q_{0+1}^{Z,U}
	\end{gathered}
\end{equation}
(for the $X$-basis, one can obtain corresponding formulas by replacing $Z$ with $X$ and $\mu$ with $\nu$).

The states in Eq.\,\,\eqref{eq:logicalstatesWODS} belong to the qutrit space ${\mathcal{H}_{0+1} = \mathcal{H}_\text{0} \oplus	\mathcal{H}_\text{1}}$. While this prevents direct application of the security proof from \cite{fung2006security}, one can show that Eqs.\,\,\eqref{eq:FLrate}--\eqref{eq:phaseError} remain valid in this case. So, the secret key rate can still be expressed using Eq.\,\,\eqref{eq:keyRateDecoy}, which now takes the following form:
\begin{equation}\label{eq:keyRateWithoutDecoy}
	R = Q_{0+1}^{Z,L}\,r(E_{0+1}^{Z,U}, E_{0+1}^{X,U}) - f_\text{ec}
	Q_{\mu} h(E_{\mu})
\end{equation} 
(here, we again account for the fact that the key consists exclusively of bits from the $Z$-basis).

\begin{figure}[t]
	\includegraphics[width=\columnwidth]{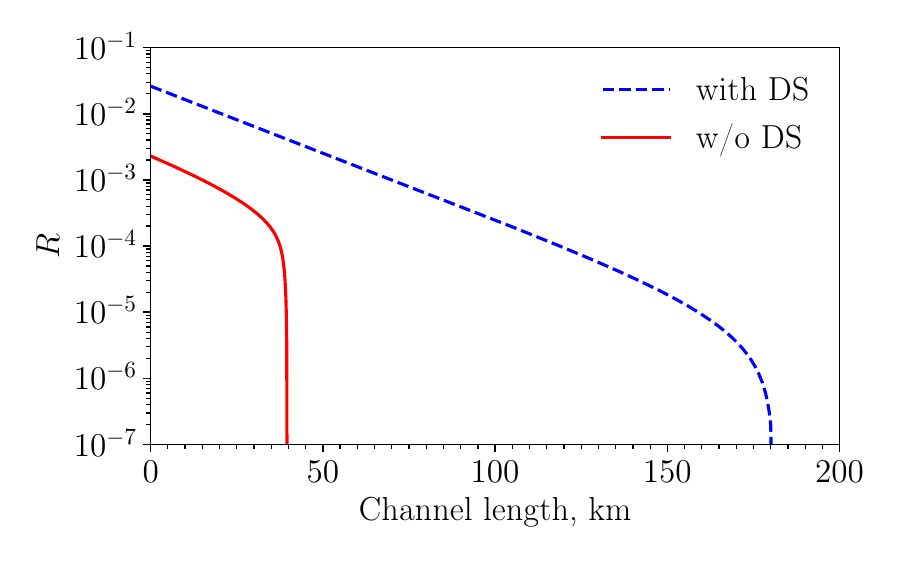}
	\caption{\label{fig:keyrate_simulation} Theoretical dependences of the secret key rate on the quantum channel length, assuming a fiber loss of 0.2\,dB/km. The dashed curve represents the three-state protocol with decoy states, while the solid line shows the corresponding protocol without decoy states.}
\end{figure}

Figure\,\,\ref{fig:keyrate_simulation} presents theoretical dependences of the secret key rate on the quantum channel length for a decoy-free three-state QKD protocol. To simulate the gain $Q_\gamma$ and bit error rate $E_\gamma$ for states with intensity $\gamma$, we used the following relations (see, e.\,g., \cite{Ma2005practicalDS}):
\begin{equation}
	\begin{split}
		Q_\gamma&=1-(1 - p_\text{dc})e^{-t\epsilon \gamma},\\
		E_\gamma&=\frac{{p_\text{dc}}/{2}+E_\text{d}(1-e^{-t\epsilon\gamma})}{1-(1
			- p_\text{dc})e^{-t\epsilon \gamma}},
	\end{split}
\end{equation}
where $p_\text{dc}$ is the dark count probability, $\epsilon$ represents the detection efficiency, and $E_\text{d}$ denotes the probability of erroneous detection. The channel transparency is given by ${t=10^{-\xi L/10}}$, where $\xi$ is the channel loss coefficient and $L$ is the fiber length. The QKD system parameters used in our simulations are summarized in Table\,\,\ref{tab:QKDsimulations}. To convert the dimensionless key rate $R$ shown in Fig.~\ref{fig:keyrate_simulation} to physical units (bits/s), it is sufficient to multiply $R$ by the quantum state preparation frequency $f$ and by the basis matching probability given by the product $p_Z^A p_Z^B$.

\begin{table}[t]
	\caption{Parameters of a QKD system used for key rate \mbox{simulations}.}\label{tab:QKDsimulations}
	\centering
		\begin{tabular}{lc}
			\hline\hline\rule{0mm}{3mm}
			\textrm{Parameter}&
			\textrm{Value}\\ \hline\rule{0mm}{3mm}
			Fiber losses $\xi$, dB/km & 0.2 \\ \rule{0mm}{3mm}
			Detector efficiency $\epsilon$ & 0.15\\ \rule{0mm}{3mm}
			Dark count probability $p_\text{dc}$& $10^{-6}$ \\ \rule{0mm}{3mm}
			Detection error probability $E_\text{d}$ & 0.01\\ \rule{0mm}{3mm}
			Error correction efficiency  $f_\text{ec}$   & 1.22 \\ \rule{0mm}{3mm}
			Basis selection probabilities  &0.5 \\ \rule{0mm}{3mm}
			Intensities without DS: &  \\
			\hspace{4em} $\nu=2\mu$ & 0.048 \\ \rule{0mm}{3mm}
			Intensities with DS: &  \\
			\hspace{4em} $\nu_0=2\mu_0$ &  1.314 \\
			\hspace{4em} $\nu_1=2\mu_1$ &  0.066 \\
			\hspace{4em} $\nu_2=\mu_2$ &  0.0 \\
			\hline \hline
		\end{tabular}\label{tab:keyrate_simulation_parameters}
\end{table}

As expected, the maximum achievable range for QKD with decoy states (DS) significantly exceeds --- in our case, by a factor of 4.5 --- the range attainable without DS. However, such extended distances are typically unnecessary for MANs, making the substantial protocol and hardware complexity required for decoy-state implementation hardly justifiable here. The family of DS-free QKD protocols, implementable using the proposed transmitter, enables secure key distribution over distances up to 40\,km with system parameters typical for practical QKD. This range generally suffices for most real-world applications. 

According to the theoretical dependence shown in Fig.~\ref{fig:keyrate_simulation}, a quantum state preparation rate of ${f=100}$\,MHz yields a secret key rate of more than $10^4$~bit/s at distances up to 30\,km.

\section{Conclusion}\label{sec:conclusion}
The proposed time-bin encoding method, implemented with pulsed optical injection, is particularly well-suited for quantum key distribution over short distances typical for metropolitan area networks. This approach enables the development of a QKD transmitter that operates without external modulators, offering two key advantages: 1) a substantially simplified design and 2) enhanced robustness against Trojan-horse attacks due to the absence of the attack object itself. With such a transmitter, a family of three-state BB84-type protocols without decoy states can be implemented, whose secrecy was briefly analyzed here.

\bibliography{modulatorFree}

\end{document}